\documentclass{article}
\usepackage{spconf,amsmath,graphicx, float}
\usepackage{acro}
\DeclareAcronym{TF}{short=TF, long=time-frequency}
\DeclareAcronym{CNN}{short=CNNs, long=Convolutional Neural Networks}
\DeclareAcronym{STFT}{short=STFT, long=short-time Fourier transform}
\DeclareAcronym{spect}{short=spect, long=Linear Spectrogram}
\DeclareAcronym{mel}{short=mel, long=mel spectrogram}
\DeclareAcronym{logmelspec}{short=logmel, long=log-mel spectrogram}
\DeclareAcronym{PCEN}{short=PCEN, long=Per-Channel Energy Normalisation}
\DeclareAcronym{STRF}{short=STRF, long=Spectro-Temporal Filters}
\DeclareAcronym{TD}{short=TD, long=Time-Domain Filter Banks}
\DeclareAcronym{LEAF}{short=LEAF, long=LEarnable Audio Frontend}
\DeclareAcronym{JSD}{short=JSD, long=Jensen-Shannon distance}
\DeclareAcronym{GAN}{short=GANs, long=Generative Adversarial Networks}
\DeclareAcronym{vad}{short=VAD, long=Voice Activity Detection}
\DeclareAcronym{sv}{short=ASV, long=Automatic Speaker Verification}
\DeclareAcronym{bsid}{short=BSID, long=Bird Species Identification}
\DeclareAcronym{plda}{short=PLDA, long=Probabilistic Linear Discriminant Analysis}
\DeclareAcronym{stam}{short=STA-VAD, long=Spectro-Temporal Attention Voice Activity Detection}
\DeclareAcronym{agc}{short=AGC, long=Automatic Gain Control}
\DeclareAcronym{drc}{short=DRC, long=Dynamic Range Compression}
\DeclareAcronym{eer}{short=EER, long=Equal Error Rate}
\DeclareAcronym{fwhm}{short=FWHM, long=Full-Width at Half Maximum}
\DeclareAcronym{auc}{short=AUC, long=Area Under Receiver Operating Characteristic Curve}

\newcounter{daggerfootnote}

\title{Learnable frontends that do not learn: \\Quantifying sensitivity to filterbank initialisation}
\name{Mark Anderson$^{\star}$ \qquad Tomi Kinnunen$^{\dagger}$ \qquad Naomi Harte$^{\star}$\thanks{Corresponding Author: andersm3@tcd.ie}}
\address{
  $^{\star}$
    SIGMEDIA Lab, School of Engineering, Trinity College Dublin, Ireland \\
  $^{\dagger}$
    School of Computing, University of Eastern Finland, Joensuu, Finland \\
}
\begin{document}
\ninept
\maketitle
\begin{abstract}
While much of modern speech and audio processing relies on deep neural networks trained using fixed audio representations, recent studies suggest great potential in acoustic frontends learnt jointly with a backend. In this study, we focus specifically on learnable filterbanks. Prior studies have reported that in frontends using learnable filterbanks initialised to a mel scale, the learned filters do not differ substantially from their initialisation. Using a Gabor-based filterbank, we investigate the sensitivity of a learnable filterbank to its initialisation using several initialisation strategies on two audio tasks: voice activity detection and bird species identification. We use the Jensen-Shannon Distance and analysis of the learned filters before and after training. We show that although performance is overall improved, the filterbanks exhibit strong sensitivity to their initialisation strategy. The limited movement from initialised values suggests that alternate optimisation strategies may allow a learnable frontend to reach better overall performance. 
\end{abstract}
\begin{keywords}
Learnable Filterbanks, LEAF, Learnable Frontend, Sensitivity, Initialisation
\end{keywords}
\section{Introduction}\label{sec:intro}
Speech tasks utilising deep learning typically use a fixed \ac{TF} representation of the audio signal, most commonly a spectrogram obtained through the \ac{STFT}. Use of the \ac{STFT} requires design choices~\cite{Harris} including the type of window function, frame length, and frame overlap. If a non-linear scaling of the frequency axis, or compression of magnitude is desired, further choices must be made (scale, number of filters, compression etc.). These \emph{hyperparameters} may be optimised for a given task and classifier via tuning, but this increases the time spent developing models.

In recent years, there have been a number of promising frontends which learn from the data directly. These \emph{learnable frontends} turn what were previously \emph{hyperparameters} to be tuned, into \emph{parameters} to be learned. Learnable frontends have the potential to provide  \ac{TF} representations of an input signal tailored to a given task or architecture, in exchange for additional computational cost during training and inference. The promise of learning the \ac{TF} representation in different downstream audio tasks ranging from speech recognition~\cite{zeghidourASR} to music tagging~\cite{musictagging} has been demonstrated. Additionally, the authors have shown similar positive findings for another class of challenging acoustic signals -- bird audio~\cite{andersonIWAENC}.

\begin{figure*}[h!]
    \centering
    \includegraphics[trim={0 2 0 3},clip, draft=false,width=0.85\textwidth]{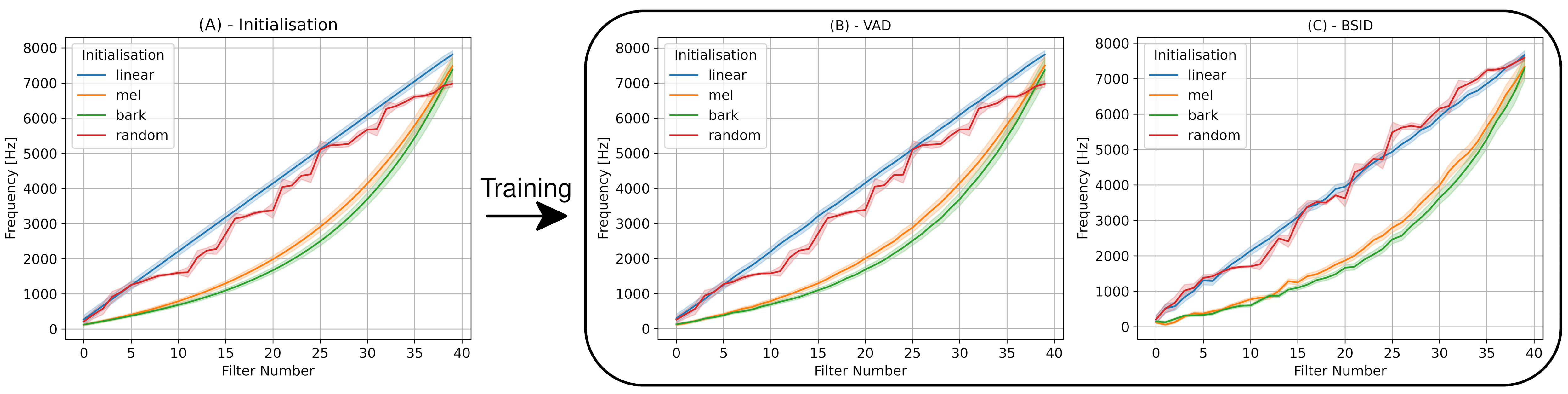}
    \caption{\small (A) shows the frequency response of each initialisation. Centre frequency is represented by the solid line and bandwidth by the shaded area. In this paper we use four initialisation types: `linear' (equally spaced, constant bandwidth), `mel' \& `bark' (psychoacoustic pitch scales) and `random' (ordered by frequency). B and C show the learned responses of the filters for the \ac{vad} and \ac{bsid} tasks after training}\label{fig:inits}
\end{figure*}

Unlike what one might intuitively expect, a recurring finding reported independently in learnable filterbank studies~\cite{TDFBanks,LEAF,EfficientLeaf, CakirFilterbank} is that the learned filters do not differ substantially from their initialised values (see Fig.~\ref{fig:inits}). These frontends are typically initialised to a mel scale and evaluated on speech tasks such as keyword spotting, speaker identification and emotion recognition \cite{LEAF, EfficientLeaf}. It is tempting to say that if the learned filters do not differ substantially from their initialisation (e.g. mel) then the initialisation was already well-suited for the task (e.g. speech recognition). While this may be true for some tasks, it does not explain the same phenomena occurring when using different filterbank initialisations. We feel this is more likely an optimisation problem, similar to the EM algorithms sensitivity to initial values~\cite{BIERNACKI2003561, EMmelynkov} and subsequent convergence to local optima.

Motivated by the above, we seek to answer the following question: are modern learnable frontends with trainable filterbanks sensitive to filterbank initialisation? To this end, our set-up involves two different audio tasks, namely \ac{vad} and \ac{bsid}. The two tasks and domains serve to complement each other: since the frequency range of bird vocalizations ($\sim$ 800Hz -- 8KHz) differs drastically from that of human speech (most energy concentrated around 300Hz -- 3.4KHz), we expect differences in filterbank learning. We consider four initialisation strategies in both tasks, including an intentionally sub-optimal \emph{random} initialisation for reference purposes. For each task, we evaluate the difference between the initialised and final filters.

Prior studies~\cite{TDFBanks, LEAF, EfficientLeaf, CakirFilterbank} have been dismissive of the lack of learning in filterbank frontends. However, as in any optimisation problem, an investigation of the sensitivity to initialisation is crucial in avoiding sub-optimal solutions. We propose to quantify the lack of learning with the \acl{JSD} and analysis of the final filterbanks. The main novelty of our study is this detailed quantification and interpretation of the difference between initialised and final filters for multiple initialisation strategies, and the determination of whether each initialisation leads to a local optima.
%Our aim is to showcase that initialisation bias is present in all initialisation regimes. 

% In Section~\ref{sec:frontends} we discuss learnable frontends in general and detail our chosen frontend: \ac{LEAF}. In Section~\ref{sec:init} we discuss filter initialisations, and initialisation bias. In Sections~\ref{sec:experiments} and~\ref{sec:results} we present our experiments, their results and accompanying discussion. Section~\ref{sec:conc} concludes the paper.

\section{Learnable Frontends}\label{sec:frontends}
Learnable audio frontends encompass a wide range of functionality, including filterbanks~\cite{TDFBanks, LEAF}, temporal downsamplers~\cite{LEAF}, and magnitude compressors~\cite{pcen_original} learnt from data. Some frontends specialise in one particular aspect, whereas others combine these elements in one pipeline, such as the recently proposed \ac{LEAF}~\cite{LEAF}.  The authors in~\cite{EfficientLeaf} categorise frontends implementing learnable filterbanks based on two criteria: (1) the domain of operation (time or frequency domain); and (2) whether filter responses are learned directly or via a parameterised function. 

For (1), parametric filters in the frequency domain need few parameters (e.g. centre frequency, bandwidth)~\cite{CakirFilterbank,FBEM,FastAudio,data-driven-fb}, but computation requires the \ac{STFT} and the additional design choices of window function and framing settings (which we wish to avoid). Time-domain filterbanks have become more common~\cite{TDFBanks, LEAF, SincNet} and are highly suitable for use with CNNs~\cite{palaz13_interspeech}. Regarding (2), learning coefficients directly~\cite{TDFBanks} offers the most freedom and may lead to better performance~\cite{andersonIWAENC}, but this increases the number of parameters and training time of a model. Coefficients generated via a parameterised function~\cite{LEAF,SincNet} not only reduce the number of parameters, but assign meaning to those parameters (typically relating to centre frequency and bandwidth). Although performance may degrade slightly, fewer and directly meaningful parameters can be beneficial, providing a path to explainability.

Regardless of form, filters are usually initialised based on a static filterbank. Since many of these frontends are aimed at human speech tasks, the mel scale is most often used. We discuss other initialisation regimes used in this paper below in Section~\ref{sec:experiments}. The frontend chosen for analysis in this work is \ac{LEAF}~\cite{LEAF} due to its multiple trainable sub-components covering the relevant aspects of frontends. In particular \ac{LEAF} is chosen for its parameterised time-domain filterbanks. \ac{LEAF} utilises learnable filters, as well as learnable low-pass filtering for temporal downsampling and learnable compression. We provide a brief overview of its operation below.

\subsection{\acf{LEAF}}
\ac{LEAF}~\cite{LEAF} is a learnable frontend consisting of three learnable layers, and one fixed operation. In this paper, we use an improvement on \ac{LEAF}, known as EfficientLEAF~\cite{EfficientLeaf} for the filterbank and temporal downsampling sections. We do not use EfficientLEAF's proposed learnable compression method, opting to use \ac{PCEN} (as in the original LEAF implementation \cite{LEAF}) instead as its effects and usage~\cite{andersonIWAENC, pcen_original, PCENwhy} is well documented.

The operation of \ac{LEAF} is as follows. The input audio signal is convolved with a set of band-pass filters (Eq.~\ref{eqn:leaf_gabor}) of length $W$ ($|t| \leq \frac{W}{2}$) in the time-domain. These filters are parameterised by the centre frequency ($\eta_n \in [0, 1]$) and bandwidth ($\sigma_{n_{bw}} \in (0, \frac{F_s}{W + 1}]$). The output of the band-pass filtering operations produce time sequences at the same temporal resolution as the input. To ensure the filterbank is analytic (i.e. frequency response contains no negative frequency components), the squared modulus of these time sequences are calculated.
\vspace{-.5em}
\begin{align}
  \phi_n(t) &= e^{2\pi j\eta_n t}\frac{1}{\sqrt{2\pi}\sigma_{n_{bw}}}e^{-\frac{t^2}{2\sigma^2_{n_{bw}}}} \label{eqn:leaf_gabor} \\
  \Phi_n(t) &= \frac{1}{\sqrt{2\pi}\sigma_{n_{lp}}}e^{-\frac{t^2}{2\sigma^2_{n_{lp}}}} \label{eqn:leaf_lowpass}
  % \text{for } &t \in \big[\frac{-W}{2}, \ldots , \frac{W}{2}\big] \nonumber
\end{align}
\vspace{-.5em}

To downsample in time, \ac{LEAF} convolves these time sequences with a learnable low-pass Gaussian filter (one per frequency band, see Eq.~\ref{eqn:leaf_lowpass}) parameterised by its bandwidth $\sigma_{n_{lp}}$. This is followed by subsampling, achieved practically via strided convolution, producing a \ac{TF} representation.

The final stage in the \ac{LEAF} pipeline is a learnable \ac{PCEN} layer, which consists of an \emph{\acf{agc}} and \emph{\acf{drc}} parameters. The \ac{agc} is applied prior to \ac{drc} and yields Eq.~\ref{eqn:pcen}. Both the \ac{agc} and \ac{drc} are learned per frequency channel.
\vspace{-.25em}
\begin{align}
    \text{PCEN}(t,f) = \left(\frac{E(t,f)}{{(M(t,f) + \epsilon)}^{\alpha}} + \delta\right)^{r} - \delta^{r} \label{eqn:pcen}
\end{align}
The \ac{agc} is implemented using a learnable smoother (Eq.~\ref{eqn:pcen_smoother}). The division of the input \ac{TF} representation ($E(t,f)$) by the smoothed \ac{TF} representation ($M(t,f)$), emphasises changes relative to the recent spectral history along the temporal axis~\cite{PCENwhy}.
\begin{align}
    M(t,f) = (1 - s)M(t-1,f) + sE(t,f) \label{eqn:pcen_smoother}
\end{align}
The \ac{drc} section is controlled by $\delta$ and $r$, where $\delta$ roughly corresponds to the threshold parameter, and $r$ the compression ratio. Higher values of $r$ correspond to less compression.

\begin{table*}[h]
\centering
\scriptsize
\setlength\tabcolsep{8pt}
\begin{tabular}{|lll|lll|}
\hline
\multicolumn{3}{|c|}{\textbf{Voice Activity Detection (VAD)}} & \multicolumn{3}{c|}{\textbf{Bird Species Identification (BSID)}} \\ \hline
\textbf{Dataset} & \textbf{Name} & TIMIT~\cite{timit} & \textbf{Dataset} & \textbf{Name} & BirdCLEF2021~\cite{birdclef2021} \\
\textbf{} & \textbf{Sample Rate} & 16KHz &  & \textbf{Sample Rate} & 16KHz (Resampled) \\
 & \textbf{No. Speakers (Train)} & 630 &  & \textbf{No. Species} & 397 \\
 & \textbf{No. Speakers (Test)} & 168 &  & \textbf{} &  \\
 & \textbf{Sentences per speaker} & 10 &  & \textbf{Recordings per species} & Variable \\
 & \textbf{Test Set} & Dedicated &  & \textbf{Test Set} & Hold-out (15\%) \\
 & \textbf{Normalisation} & $-6$dbFS &  & \textbf{Normalisation} & $-6$dbFS \\ \hline
\textbf{Classifier} & \textbf{Model} & STA-VAD~\cite{STAM} & \textbf{Classifier} & \textbf{Model} & EfficientNet-B0~\cite{EfficientNet} \\
 & \textbf{Initial Learning Rate} & 0.001 &  & \textbf{Initial Learning Rate} & 0.001 \\
 & \textbf{Optimiser / Scheduler} & ADAM / Cosine Annealing &  & \textbf{Optimiser / Scheduler} & ADAM / Cyclic \\
 & \textbf{Metrics} & F1, AUC & \textbf{} & \textbf{Metrics} & F1, Acc. \\ \hline
\end{tabular}
\caption{\small Details of dataset and classifier for the \ac{vad} and \ac{bsid} tasks}
\label{tab:tasks}
\end{table*}

\section{Experimental Setup}\label{sec:experiments}
\subsection{Frontend Initialisation}
In this work, we use four different initialisation strategies. Two are based on psychoacoustic scales, \emph{mel}~\cite{mel} and \emph{bark}~\cite{bark}. The other two initialisations are labelled \emph{linear} and \emph{random}. Both \emph{mel} and \emph{bark} are well suited to human speech~\cite{mel,bark,melhuman}. Prior work on bird audio offers no consensus~\cite{Stowell2022} on the usage of \emph{mel} or \emph{bark} over other scale --- some favouring \emph{mel}~\cite{bird_mel} but others finding it sub-optimal for bird audio~\cite{FBEM}. The \emph{linear} initialisation is seen as sub-optimal for both tasks~\cite{bird_mel, optimalmapping} and \emph{random} is purposefully designed to be sub-optimal on both tasks

For \emph{mel} and \emph{bark}, centre frequencies are linearly spaced in the respective scales. The bandwidth parameter $\sigma_{n_{bw}}$ is set to match the \emph{\ac{fwhm}} of an equivalent triangular filter. This corresponds to the $-3$dB point. The \emph{linear} initialisation has linearly spaced centre frequencies. Bandwidth is constant per filter, with $\sigma_{n_{bw}}$ set in the same manner as above.  The \emph{random} initialisation is achieved by uniform sampling of valid frequency values to determine the centre frequencies of the filters. In order to cover the entire desired frequency spectrum, these centre frequencies are sorted, and $\sigma_{n_{bw}}$ is set such that the bandwidth covers at least the \ac{fwhm} of the filters either side. For reproducibility the same seed is used in all relevant experiments. Fig.~\ref{fig:inits}(A) shows the resulting filterbank from each initialisation strategy in the frequency domain.

\subsection{Experiments}
We evaluate on two audio tasks, \acf{vad} and \acf{bsid}. In all experiments we use 40 filters. Additionally, we train fixed filterbank models of each initialisation using \ac{LEAF}. In the fixed filterbanks experiments the \ac{PCEN} layer is still trainable; we only wish to fix the filterbank for analysis purposes. As a baseline, we use Log-Mel spectrograms with 40 mel filters (Static Log-Mel). Table~\ref{tab:tasks} summarises both experiments.

\textbf{\acl{vad}:} In the \ac{vad} task, we use the TIMIT corpus~\cite{timit} and \ac{stam}~\cite{STAM}. TIMIT provides a `clean' corpus which disjoint train/test speakers, which is easily augmented with additive noise. \ac{stam} is a recently proposed model which is noise-robust and lightweight. Similar to~\cite{STAM}, we mitigate the speech/non-speech class imbalance through additional silence before/after each utterance. The length of this silence is randomly chosen between 0.5--1s. The corpus is augmented with additive noise (between $-10$dB and $30$dB) from the MSNS dataset~\cite{reddy2019scalable}, which contains fourteen noise types (including stationary and non-stationary noise). The noise dataset also contains test data (same noise categories, different recordings). During training, SNR is adaptive, through a callback which increases the maximum allowed SNR by $5dB$ (starting from $-10$dB) when val-loss plateaus. Each epoch involves shuffling the data, and an equal split of data to be augmented with randomly selected noise at the available SNR values. The power of the added noise is calculated based only on segments containing speech. 

The model was trained using a 90:10 train/val split. We utilise a Cosine Annealing learning rate scheduler~\cite{cosineannealing}, set to restart approximately when new SNR values are introduced. Noise levels in validation stages are a random selection of the currently available SNR values. Test data was taken from the TIMIT test set and augmented with additional silence and additive noise at a SNR of $15$dB. Additive noise in the test set is randomly selected from all noise categories. In line with \cite{LEAF, EfficientLeaf, STAM} we report accuracy and \ac{auc} in our experiments.

\textbf{\acl{bsid}:} In the \ac{bsid} task, we use the publicly available BirdCLEF2021~\cite{birdclef2021} training dataset and EfficientNet-B0~\cite{EfficientNet}. Both dataset and model were used to evaluate EfficientLEAF \cite{EfficientLeaf}. EfficientNet-B0 is also used extensively to evaluate the original implementation of \ac{LEAF}~\cite{LEAF}. The dataset contains variable length, high-quality, focal recordings taken from Xeno-Canto, an online collection of crowdsourced bird recordings. Recordings are normalised to $-6$dBFS. There are 397 species present in the dataset, from across North \& South America. There is a significant class imbalance; 12 species contain 500 recordings, 9 species contain less than 25 recordings. We utilise a 70:15:15 split of this dataset (accounting for class imbalance) for training, evaluation and testing respectively. We report both accuracy (used to evaluate EfficientLEAF~\cite{EfficientLeaf}) and F1-Score (official metric of BirdCLEF2021).

\subsection{Evaluation of Frontend Sensitivity}
Evaluation of sensitivity due to initialisation requires an adequate metric to quantify the difference between the initialised and final filterbanks. We use the \acf{JSD}. The \ac{JSD} (Eq.~\ref{eqn:JSD}) is a means of measuring the difference of two probability distributions. It takes the form of a symmetrical Kullbeck-Leibler divergence (Eq.~\ref{eqn:KLD}), comparing each distribution $P$ and $Q$ to a mixture distribution $M$. The \ac{JSD} is a true metric~\cite{JSDMetric} that satisfies all related axioms (including the triangle inequality), and is bounded by $[0, 1]$ when a logarithmic base of 2 is used in Eq.~\ref{eqn:KLD}.
\vspace{-.25em}
\begin{align}
  D_{\mathrm{JS}}(P, Q) &= \sqrt{\frac{1}{2}\big[D_{\mathrm{KL}}(P||M) + D_{\mathrm{KL}}(Q||M)\big]} \label{eqn:JSD}\\
  \text{Where, } M &= \frac{P + Q}{2} \nonumber\\
  \text{And, } D_{\mathrm{KL}}(A||B) &= \sum_{x}A(x)\log\left(\frac{A(x)}{B(x)}\right) \label{eqn:KLD}
\end{align}
The frequency response of each filter in \ac{LEAF} is a sampled Gaussian kernel and can be interpreted similarly to a probability distribution. Typically used to compare between a ground truth distribution and a distribution of simulated values, for our purposes it compares between an initial distribution and a final distribution. 

\begin{table}[]
\begin{center}
\footnotesize
\setlength\tabcolsep{3pt}
\begin{tabular}{l|cccc|cccc|}
\cline{2-9}
 & \multicolumn{4}{c|}{\textbf{VAD}} & \multicolumn{4}{c|}{\textbf{BSID}} \\ \cline{2-9} 
\textbf{} & \multicolumn{2}{c|}{\textbf{F1-Score}} & \multicolumn{2}{c|}{\textbf{AUC}} & \multicolumn{2}{c|}{\textbf{F1-Score}} & \multicolumn{2}{c|}{\textbf{Acc (\%)}} \\ \hline
\multicolumn{1}{|l|}{\textbf{Filterbank}} & \textit{Fixed} & \multicolumn{1}{c|}{\textit{Learn}} & \textit{Fixed} & \textit{Learn} & \textit{Fixed} & \multicolumn{1}{c|}{\textit{Learn}} & \textit{Fixed} & \textit{Learn} \\ \hline
\multicolumn{1}{|l|}{Linear} & 0.841 & \multicolumn{1}{c|}{\textbf{0.862}} & 0.839 & \textbf{0.858} & 0.663 & \multicolumn{1}{c|}{\textbf{0.674}} & 71.6 & \textbf{73.4} \\
\multicolumn{1}{|l|}{Mel} & 0.826 & \multicolumn{1}{c|}{0.858} & 0.778 & 0.840 & 0.660 & \multicolumn{1}{c|}{0.668} & 71.4 & 71.8 \\
\multicolumn{1}{|l|}{Bark} & 0.834 & \multicolumn{1}{c|}{0.860} & 0.816 & 0.846 & 0.660 & \multicolumn{1}{c|}{0.665} & 70.8 & 71.6 \\
\multicolumn{1}{|l|}{Random} & 0.802 & \multicolumn{1}{c|}{0.807} & 0.765 & 0.759 & 0.653 & \multicolumn{1}{c|}{0.662} & 68.1 & 71.5 \\ \hline
\multicolumn{1}{|l|}{Log-Mel} & 0.847 & \multicolumn{1}{c|}{-} & 0.881 & - & 0.646 & \multicolumn{1}{c|}{-} & 69.7 & - \\ \hline
\end{tabular}
\end{center}
\caption{\small Results on hold-out test set for VAD and BSID tasks. Includes results using learnable frontends with fixed filterbanks (\textit{Fixed}) and using learnable frontends with learnable filterbanks (\textit{Learn}). Results between each \textit{fixed}/\textit{learn} pair are statistically significant. Best results marked in \textbf{bold}.}\label{tab:results}
\end{table}

\begin{figure*}[h!]
    \centering
    \includegraphics[trim={0 5.5 0 1.75},clip, draft=false, width=0.795\textwidth]{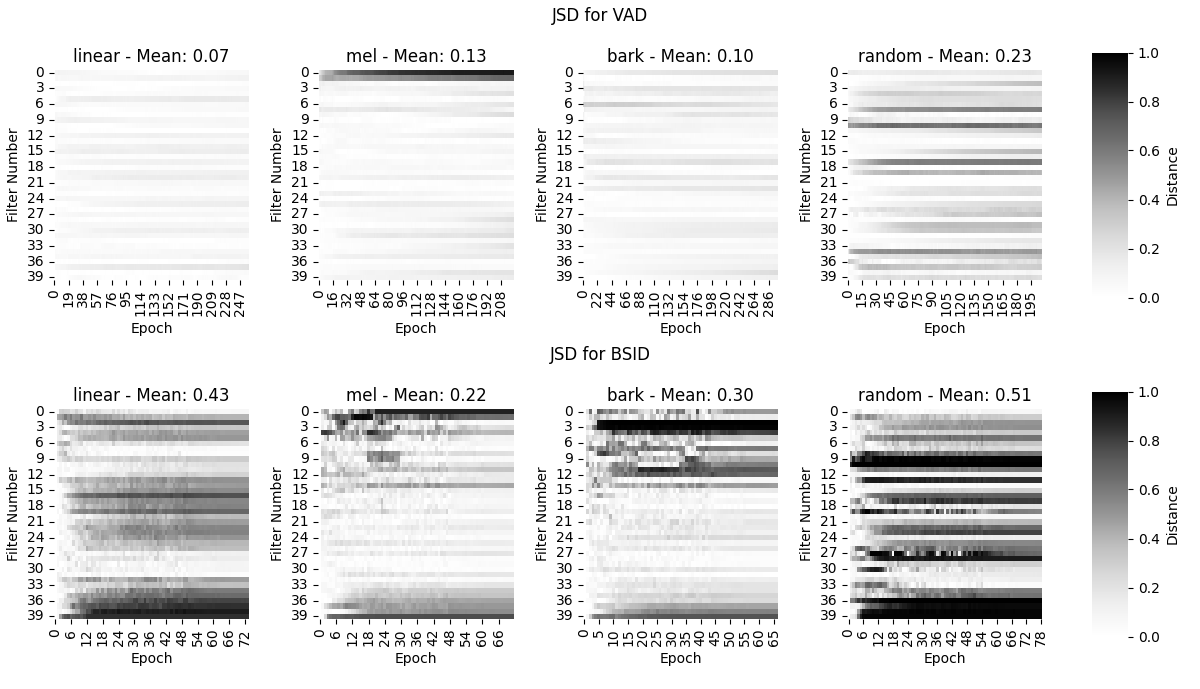}
    \caption{\small Jensen-Shannon distance of each filter from its initialisation for the VAD and BSID tasks, by initialisation strategy. The mean of the final distances is also shown in each plot's title.}\label{fig:jsd_time}
\end{figure*}

\section{Results \& Discussion}\label{sec:results}
Table~\ref{tab:results} reports performance for each task on their respective test set. These results are not the main findings of this work, but they demonstrate that each model has trained correctly. We observe that fully learnable frontends outperform their fixed counterparts; The fixed filterbank learnable frontends outperform the log-mel spectrogram baseline features with the exception of the \emph{random} initialisation (expected as the \emph{random} initialisation is intentionally sub-optimal). Although improving over their fixed counterparts, learnable filterbanks with \emph{random} initialisation perform worse than learned filterbanks using other initialisation strategies; lack of learning negatively impacts performance. Increases in performance from static baseline features to learnable features is most evident on the \ac{bsid} task. This is likely due to the \ac{PCEN} compression layer~\cite{andersonIWAENC}. Performance improvement when using learnable filterbanks is greater in the \ac{vad} task. Prior work~\cite{bird_mel, optimalmapping} declares a fixed \emph{linear} filterbank as sub-optimal on both tasks, however in both fixed and learnable cases, \emph{linear} initialisation performs best on \ac{vad} and \ac{bsid}. 

For \ac{vad}, the best result was achieved using trainable filterbanks with linear initialisation (F1 $0.862$, AUC $0.858$). Whilst a direct comparison is problematic due to a difference in testing dataset, our performance is in line with that reported in~\cite{STAM} (Mean AUC $0.886$). In the \ac{bsid} task, again best results are achieved using trainable filterbanks with linear initialisation (F1 $0.674$, Acc. $73.4\%$). Direct comparison of accuracy is possible, with our performance being in line with~\cite{EfficientLeaf} who reported an accuracy of $72.2\%$ on this task.

The more pertinent findings of this study can be seen in Figs. \ref{fig:inits} and \ref{fig:jsd_time}. Fig.~\ref{fig:inits} shows the frequency response of each filterbank before and after training. Fig.~\ref{fig:jsd_time} depicts the filterbank movement from initial values over time, using the \ac{JSD}. In Fig.~\ref{fig:jsd_time},  for the \ac{vad} task there is very little movement from the \emph{linear} and \emph{bark} initialisations ($0.07$ and $0.10$ respectively), with no filters moving substantially. The \emph{mel} initialisation ($0.13$) has movement in the lower frequency filters; there is a shift to lower centre frequencies while the bandwidth remains the same (this can be seen in Fig.~\ref{fig:inits}, although the change is subtle). This shift results in a large distance due to the small bandwidth of the low frequency filters; any change in centre frequency for low bandwidth filters represents a large change in distance. There is more movement in the intentionally sub-optimal \emph{random} initialisation, with mean distance of $0.23$. Despite greater movement than the learned \emph{linear} filterbanks, other initialisation strategies do not provide as good a solution. Contrasting these distances with the final learned filters in Fig.~\ref{fig:inits} however, we see the overall state of the filterbanks have not changed substantially from their initialisation.

Comparing these findings with the \ac{bsid} task, in Fig.~\ref{fig:jsd_time} we see more movement between initial and final filters compared to the \ac{vad} task. The \emph{mel} and \emph{bark} initialisations move least ($0.22$ and $0.30$ respectively), both showing movement in the lower and higher frequency filters. The \emph{linear} initialisation ($0.43$) has more movement than either of the psychoacoustic initialisations, with movement across most frequency channels. Similar to the \ac{vad} task, the \emph{random} initialisation shows the most movement ($0.51$). Contrasting again to Fig.~\ref{fig:inits} however, we do not see change in the overall state of the learned filterbanks with the exception of \emph{random}, which moves towards a linear-like state. Similar to \ac{vad}, although the learned filterbanks in the \ac{bsid} task exhibit some movement they do not achieve similar performance to the best performing filterbank (\emph{linear} initialisation). If the optimisation strategy was functioning as intended, all learned filters would achieve similar performance.

%Although learnable frontends with trainable filterbanks outperform their fixed counterparts and the static baseline features in terms of model performance (Table~\ref{tab:results}), the resulting filters do not differ substantially from their initialisation in most cases. Each model finds a local optima for each filterbank initialisation; Therefore learnable filterbanks are sensitive to their initialisation.

\section{Conclusion}\label{sec:conc}
In this paper we demonstrated the sensitivity of learnable filterbanks to initialisation, and quantified the change from initialisation to final learned filters. The lack of learning in learnable filterbanks was consistent across two different tasks and four initialisation strategies. Performance was improved with learnable filterbanks, but at additional computational cost

Our results have implications in considering the trade-off between training time and model utility. For learnable filterbanks to be merited, they must offer reliable performance increases. Learnable filterbanks should move from their initialisation, to a family of optimal filters. The inconsistency in the performance of the learned filters, coupled with the lack of movement from initial filterbank values, demonstrates a shortcoming in the overall optimisation strategy. 

%% Choose between lines 245 and 246
%% if 246, change fig width for JSD to 0.83 
The authors believe that our methodology and results provide novel insights into quantifying the shortcomings of learnable filterbank based audio frontends. Explaining the root causes and developing tangible mitigation strategies to these shortcomings remain important longer-term goals. The authors are currently working towards these goals.
%Further work is required to isolate the issue, but we believe that it may be due to learning rate and that separate optimisation strategies for the frontend and classifier may be required. Work is ongoing on this aspect.

\clearpage
% \let\oldthebibliography=\thebibliography
% \let\endoldthebibliography=\endthebibliography
% \renewenvironment{thebibliography}[1]{%
%    \begin{oldthebibliography}{#1}%
%      \setlength{\itemsep}{+.10ex}%
% }%
% {%
%    \end{oldthebibliography}%
% }
\bibliographystyle{IEEEbib}
\bibliography{refs}

\end{document}